\documentclass[preprint,12pt]{elsarticle}

\usepackage{graphicx}

\usepackage{amssymb}

\journal{Journal of Alloys and Compounds}

\begin{document}

\begin{frontmatter}



\title{Transport and magnetotransport properties of cold-pressed
CrO$_2$ powder, prepared by hydrothermal synthesis}


\author[label1]{B. I. Belevtsev\corref{cor1}}
\ead{belevtsev@ilt.kharkov.ua}

\author[label1]{N. V. Dalakova}

\author[label2]{M. G. Osmolowsky}

\author[label1]{E. Yu. Beliayev}

\author[label2]{A. A. Selutin}

\address[label1]{B. Verkin Institute
for Low Temperature Physics and Engineering, National Academy of
Sciences, Kharkov 61103, Ukraine}

\address[label2]{St. Petersburg State University, Department of
Chemistry, St. Petersburg 198504, Russia}

\cortext[cor1]{Corresponding author at: B. Verkin Institute for
Low Temperature Physics and Engineering, National Academy of
Sciences, Kharkov 61103, Ukraine. Fax:++38-057-3450593;
Tel.:++38-057-3410963}

\begin{abstract}
Submicron powder of CrO$_2$ was prepared by hydrothermal synthesis
method from chromium trioxide with use of special modifiers, which
govern the growth of particles. Particles obtained were of rounded
form with mean diameter about 120 nm. The powder (stabilized with
thin surface layer of $\beta$-CrOOH) has been characterized by
structural, X-ray and magnetic measurements. The powder under
investigation (with Curie temperature about 385 K) was
cold-pressed and its transport and magnetotransport properties
have been measured in the temperature range 4--450~K in magnetic
field up to 1.6 T. The samples studied is characterized by
non-metallic temperature behavior of resistance and large negative
magnetoresistance (MR) in low temperature range. At $T=5$~K the MR
magnitude has been -17\% at $H=0.3$ T and -20\% at $H=1.4$ T. Its
magnitude decreased fast with increase in temperature reducing to
0.3\% and less for $T>200$~K. It is shown that this MR behavior is
inherent for a system of magnetic grains with spin-dependent
intergrain tunnelling. Some peculiarities of MR behavior in
low-temperature range (below 40 K) can be associated with
percolating character of tunnelling conductivity of this granular
system under conditions of availability of only few conducting
current paths through the sample.
\end{abstract}


\end{frontmatter}


\section{Introduction}
\label{} Chromium dioxide (CrO$_2$) is ferromagnet ($T_C\approx
390$~K) which as a fine-particle medium has long been in use in
the magnetic recording industry. This compound is predicted
theoretically \cite{schwarz} to be a half-metal, that found
experimental support \cite{coy1,coy2,ziese}. In a half-metal,
conductivity band at the Fermi level has carriers of only one spin
type. Generally, spin polarization $P$ is determined as
$P=(N_{\uparrow}-N_{\downarrow})/(N_{\uparrow}+N_{\downarrow})$
where  $N_{\uparrow}$ and $N_{\downarrow}$ are densities of states
for the up- and down-spin electrons, respectively. For
half-metals,  $P$ is expected to be equal 100\% at low enough
temperature what gives the opportunity for spintronics and
nanotechnology devices.
\par
Intrinsic magnetoresistance (MR) of crystal CrO$_2$ is known to be
rather low (about 1\%/T at room temperature) \cite{coy2,ziese}. At
the same time, MR of the CrO$_2$ pressed powder with rather weak
links between ferromagnetic (FM) grains can reach substantial
magnitude (30\% or more) at low temperature \cite{coy1,coy3}. This
MR is extrinsic, that is determined by employed technology  of
preparation of granular material with insulating thin layers
between FM grains. Insulating interlayers precludes direct FM
exchange interaction between neighboring grains permitting,
however, electron tunnelling. This tunnelling depends on the
relative orientation of the magnetization in adjacent grains and
is very sensitive to applied magnetic field. Tunnelling
probability is maximal when moments in neighboring grains are
aligned \cite{ziese,tsymb}. MR of this type is called tunnelling
MR (TMR) with MR mechanism being, so called, spin-dependent
tunnelling \cite{ziese,tsymb}. Significant spin polarization is
the necessary prerequisite for high TMR.
\par
Previous MR studies of the CrO$_2$ compacted powders were carried
out mainly for commercial powders used for magnetic recording,
which consists of acicular particles
\cite{coy1,coy2,coy3,dai,liu}. The powders were coated with thin
surface layer of antiferromagnetic Cr$_2$O$_3$, which provides
tunnelling contact between particles and, therefore, high enough
TMR. It is of interest to study MR properties of CrO$_2$ powders
prepared by alternative procedures and coated with insulating
surface layers of different type. In this study, CrO$_2$ powder
was prepared with hydrothermal synthesis, and CrO$_2$ particles
were stabilized by converting surface layer of material into
$\beta$-CrOOH. The cold-pressed samples studied consist of rounded
particles of CrO$_2$. This ensures lesser shape anisotropy and
porosity of samples. Results of measurement of transport and
magnetotransport properties of the CrO$_2$ pressed powder are
presented below.

\section{Preparation and characterization of the chromium dioxide powder}
In this work, the synthesis of CrO$_2$ has been carried out by
hydrothermal synthesis method from mixture of 250 g of high-purity
chromium trioxide (CrO$_3$), 50 g of distilled water and some
special modifiers, which govern the growth of particles
\cite{osm}. The synthesis was performed in an autoclave from
stainless steel. Reagents have been put into test-tube of SIMAX
glass. Maximum temperature of synthesis has been 320$^{\circ}$C
under pressure of 32 MPa. The resulting compound was dried in air
at 150$^{\circ}$C and milled in mill (MRP-2) with rotating knifes.
The powder obtained has been stabilized (by converting surface
layer of material into rhombohedral $\beta$-CrOOH, chromium
oxyhydroxide) in one liter of 0.3~M neutral solution of sodium
sulfite under continuous stirring during 30 min. After that the
magnetic powder has been extracted from stabilizing solution,
washed with 2 liters of distilled water up to decolorization of
the rinsing water. The paste obtained has been dried during 2
hours in desiccator at 150$^{\circ}$C. The dried substance has
been milled again.
\par
The stabilized powder has specific surface equal to 10.5 m$^{2}$/g
and consisted of closely sized rounded polycrystalline particles
with mean diameter about 120 nm. The mean thickness of the
stabilization surface layer of $\beta$-CrOOH is about 1 nm. This
layer is insulating and non-ferromagnetic. Electron micrograph of
the CrO$_2$ powder, obtained in transmission electron microscope
JEM-100C, is shown in Fig. 1. In addition, the powder has been
characterized with the X-ray and magnetization studies. Lattice
parameters obtained are $a = 0.4419$ nm and $c = 0.2914$ nm in
rutile-like lattice that agrees well with known data for pure
CrO$_2$ \cite{coy2}. The admixture components and $\beta$-CrOOH
cause only slight shouldering of diffraction lines on the
small-angle side, since admixtures have rutile-like structure and
$\beta$-CrOOH is formed topotactically from CrO$_2$. Due to slight
lattice mismatch $\beta$-CrOOH is somewhat distorted. The
stabilizing $\beta$-CrOOH layer in our synthesis method has
orthorhombic lattice, same as was found in Ref. \cite{essig}.
\par
The powder obtained was pressed in a hydraulic press at 5 MPa to
tablets with dimensions $3\times 5\times 10$ mm$^3$. Density of
tablets consists 60\% of the full X-ray density. Magnetic
properties were measured in vibrating sample magnetometer (77 Hz).
Examples of temperature dependences of the magnetization are shown
in Fig. 2. The Curie temperature, $T_C$, was found to be
$T_C\approx 112^{\circ}$C (385 K). At room temperature the
compacted powder has coercive force about 0.0149 T, specific
magnetization at $H=1$~T about 62.5 Am$^2$/kg, remanent
magnetization $M_r\approx 14.2$~Am$^2$/kg. At liquid nitrogen
temperature (77 K), magnetization at 1 T increases up to 110.5
Am$^2$/kg, which can be taken as a low estimate for saturation
magnetization $M_s$ (at liquid-helium temperatures range the
specific magnetization should be evidently somewhat greater). It
is known \cite{coy2} that $M_s\approx 133$~Am$^2$/kg for pure
CrO$_2$ that corresponds to magnetic moment per formula unit about
2 $\mu_{B}$. This value is larger then that measured in this work,
which can be partly determined by the stabilization interlayers of
non-magnetic $\beta$-CrOOH in the sample studied. Porosity
determines internal demagnetization fields \cite{smit}, that leads
to rather low ratio $M_r/M_s$ in the sample studied, which is
about 0.23  at room temperature taking the above-mentioned
magnitude of magnetization at $H=1$~T as $M_s$.

\section{Transport and magnetotransport properties}
\subsection{General characterization}
Transport and magnetotrasport studies have been performed for two
tablets obtained as indicated above. It is found that both tablets
have essentially equal properties.  Resistance as a function of
temperature and magnetic field (up to 1.6 T) was measured using a
standard four-point probe technique in a home-made cryostat.
Current-voltage characteristics were linear for current below 0.2
mA. The measurements have been carried out in this low-current
range where Ohm's law is obeyed. Temperature curves for the
resistivity are shown in Fig. 3. The curve 1 was recorded on
heating of the as-prepared sample from 5 K to 430 K (after
preliminary cooling down to 5 K in zero field). Resistance has
strong non-metallic behavior ($dR/dT<0$) and changes by nearly two
order of magnitude in this temperature range. Nevertheless, $R(T)$
dependence is close to exponential one only at low temperature
($T<20$~K). Narrowness of the temperature range does not permit us
to determine exact form of this dependence.
\par
Non-metallic behavior of $\rho(T)$ and high value of $\rho$ (as
compared with that of pure CrO$_2$ for which according to Ref.
\cite {coy3} residual resistivity is about $10^{-5}$~$\Omega$~cm)
indicate that resistivity is determined by tunnelling of charge
carriers between grains of CrO$_2$ separated by $\beta$-CrOOH
layers. In this case high TMR can be expected and is found.
Magnetoresistance, $[R(H)-R(0)]/R(0)$, measured in the field
direction perpendicular to the current has appeared quite large
(-20\% at $T=5$~K for $H=1.2$~T, see Fig. 4) that agrees with that
reported by other authors for CrO$_2$  powder compact
\cite{coy1,coy2,coy3,dai,liu}. There are reasons to believe that
such giant magnitudes of MR are determined by magnetic tunnelling
between grains with high spin polarization \cite{ziese,tsymb}.
Extrinsic MR of this type is observed also in other compounds with
high spin polarization of charge carriers, for example in
mixed-valence manganites \cite{coy1,ziese,gupta}.
\par
Specific feature of TMR is its rather rapid decline with
temperature \cite{coy1,coy3,liu,gupta}, found in this study as
well (Fig. 4). We have found that the temperature dependence of MR
of the sample at $H=1.2$~T is described well by relation
$-[R(H)-R(0)]/R(0)\propto \exp(T/T_{mr})$ where $T_{mr} \approx
45$~K (Fig. 4), in good agreement with results of other studies of
CrO$_2$  compacts \cite{coy1,coy3,liu}. This characteristic
feature of the TMR is usually connected with strong decrease in
polarization $P$ with increasing temperature
\cite{tsymb,gupta,sun}. Other possible reasons of rapid decrease
in MR of half-metallic granular systems with  increasing
temperature are considered in Refs. \cite{ziese,coy3,tsymb,liu}.
\par
During the measurements in the cryostat, the samples were kept
rather long time in vacuum ($\approx 10^{-5}$~Torr). Apparently
for this reason, the samples have appeared as unstable to heating
above room temperature. Replicated measurements of $\rho(T)$ after
heating to 430 K show  a considerable decrease in resistivity
(Fig. 3) and MR (Fig. 4). More careful study has revealed that
irreversible changes in the resistivity begin at temperature which
is not so far above room temperature. It is evident that these
changes in resistive and magnetoresistive properties are
determined by eroding of intergrain stabilization $\beta$-CrOOH
layers, and formation as well as strengthening of direct
electrical interconnections between the CrO$_2$ grains. The
eroding is evidently accelerated when samples were placed in a
rather high vacuum, since in air a significant eroding is not
expected below 150$^{\circ}$C.  A quite possible reason for this
is that in vacuum the chromic acid residual liberates water in an
accelerated way. This causes formation of CrO$_2$ and, therefore,
localized thinning of stabilizing layer and occurrence of local
short-circuiting bridges, which lead to decrease in the
resistivity and MR. Stability of the $\beta$-CrOOH layer and
mechanisms of its degradation need further research.

\subsection{Field dependences of resistance}
Magnetic-field dependences of resistance of the samples studied
reflect their magnetic and structural properties  in line with the
suggested tunnelling character of conductivity and MR. Curves
$R(H)$ have been recorded for different fixed temperatures
according to the pattern similar to that used for routine
measurements of hysteresis cycles of the magnetization. Initially,
the field was increased up to $H_{max}$ (1.3--1.6 T). Thereupon
$R(H)$ curves were taken in subsequent cycles of $H$ between $\pm
H_{max}$, as shown in Figs. 5--8  for as-prepared state of the
sample. The $R(H)$ curves were hysteretic with hysteresis
enhancing dramatically with decreasing temperature (Figs. 5--8).
For further discussion of the hysteresis phenomena we shall
distinguish curves $\Delta R(H)/R(0)$ by direction of the field
sweeping. The first type MR curves are taken with changing field
from $+H_{max}$ to $-H_{max}$; whereas, the second type curves are
recorded with field sweeping in the opposite direction, from
$-H_{max}$ to $+H_{max}$. We denoted these types of MR curves,
respectively, by MR$(\leftarrow)$ and MR$(\rightarrow)$ as shown
in Figs. 5 and 7.
\par
In low-field range, all MR curves show two not very high peaks of
positive MR, which are symmetrically positioned about $H=0$. These
peaks of approximately the same height are positioned at fields
$H=H_p$ and $H=-H_p$ (position of $H_p$ is shown in Figs. 5 and
6). With increasing temperature the peaks become rather flat (see
Fig. 8 for $T=50$~K). Such type of hysteresis is generally
expected for TMR of half metallic granular samples. It is believed
\cite{ziese,tsymb,sun} that MR curves in this case reflect
magnetization hysteresis cycles so that resistance is expected to
be maximum at demagnetized state of the sample at zero
magnetization, which takes place at $H= H_c$ or $-H_c$, where
$H_c$ is coercive force. So that the relation $H_p\approx H_c$ is
expected, which was really found for some half-metallic systems
including CrO$_2$ powder compacts \cite{ziese,coy3}. As
magnetization increases with increasing field, mutual alignment of
magnetic moments in adjacent grains becomes stronger resulting in
an increase in MR. It can be expected, therefore, that, when $M$
goes close to the saturation value $M_s$, MR will go to some
saturation limit as well. This behavior seems to be true when
sample was studied at low enough temperature (Fig.~5).
\par
For the samples studied, the simple correlation
\cite{ziese,tsymb,sun} between hysteretic behaviors of $R(H)$ and
$M(H)$ appears to be valid only for high enough temperature (Figs.
7 and 8). Here MR curves  taken for different sweep directions,
[MR$(\leftarrow)$ and MR$(\rightarrow)$], merge together at a high
enough field like corresponding hysteretic $M(H)$ curves. At low
temperature (below 15 K), however, an additional hysteresis effect
(crossing of the MR$(\leftarrow)$ and MR$(\rightarrow)$ curves) is
clearly seen for fields not far above $|H_{p}|$ (Figs. 5 and 6)
which is inconsistent with the simple model picture for a single
magnetic tunnel junction. It is no surprise that behavior of the
sample studied, which presents an array of tunnel junctions,
differs from that expected for single junction.
\par
It is known that quantum tunnelling of the charge carriers occurs
between states of equal energy. Actually, however, there is always
some energy level mismatch between electron states in neighboring
grains for different reasons. In this case, a charge carrier
should gain some energy (for example, from phonons) to accomplish
the tunnelling. The intergrain conductivity is conditioned,
therefore, by the two processes: the tunnelling and thermal
activation.
\par
In real granular metals, thickness of insulating interlayers
between grains is not the same throughout the system, thus, the
conductivity is percolating \cite{sheng}. It is determined by the
presence of ``optimal'' chains of grains with maximum probability
of tunnelling for adjacent pairs of grains forming the chain. In
conditions of activated conductivity, number of conducting chains
decreases with decreasing temperature, so that at low enough
temperature a percolation network can even come to a single
conducting path \cite{sheng}. These ``optimal'' chains have some
weak links (high-resistance tunnel junctions) with increased
activation energy, that cannot be avoided. These high-resistance
junctions, in fact, determine the activated character of total
measured conductivity.
\par
Picture of tunnelling conductivity in granular half-metallic
systems is more intricate than that in non-magnetic systems. Here
the tunnelling probability depends not only on properties of
insulating intergrain barriers but also on the mutual alignment of
magnetic moments in neighboring grains. It is clear, therefore,
that at a fixed temperature the spacial positions of the
``optimal'' chains of grains with maximum conductivity (and the
positions of high-resistance junctions within chains, which
determined the whole system resistance) are continuously changing
with variation of external magnetic field. Moreover, application
of an external magnetic field induces opening of additional
transport channels \cite{ju} that causes giant MR in such
percolating systems. The additional MR hysteresis (Figs. 5 and 6)
in the field range above $H_p$ is evidently takes place only in
conditions of small number of percolating current channels at low
enough temperature. This can indicate that varying percolating
sets of current channels are different for increasing and
decreasing magnetic field in this field range.
\par
It is generally believed \cite{sheng} that activation energy is
mainly determined by Coulomb charging energy $E_C=e^{2}/\kappa d$,
where $\kappa$ is effective dielectric constant and $d$ is average
grain diameter. Taking $\kappa=5$ for CrO$_2$ compact powder (as
that in Ref. \cite{coy3}) and $d\approx 120$~nm, a charging energy
for the sample studied is about 28 K. It is not surprising then
that, as indicated above, $R(T)$ dependence is close to
exponential one only for $T<20$~K. In any case, rising temperature
leads to increase in effective number of current paths through the
sample and, therefore, to more homogeneous current distribution
within the sample \cite{sheng,ju}. The conductivity in magnetic
field is determined in this case by magnetic state of far more
number of grains, than in the case of few current paths at low
temperature, so that hysteretic behaviors of $R(H)$ and $M(H)$ are
mutually correlated at fairly high temperature in the expected
way.
\par
Occurrence  of the additional MR hysteresis at low temperature
range (Figs. 5 and 6) is accompanied by peculiar temperature
behavior of characteristic field $H_p$ (which is point of maximum
resistance in the MR hysteresis cycle). With increasing
temperature $H_{p}(T)$ first rises and then (above 30 K) goes down
(Fig. 9). Decrease in $H_p$ with decreasing temperature below 30 K
is unexpected under common suggestion $H_p\approx H_c$.
\par
At first glance the sample studied can be considered as a system
of single-domain grains with weak exchange coupling between the
grains. Really, the critical single-domain diameter of spherical
particles for CrO$_2$ is estimated to be about 200 nm \cite{kron}
which is larger than average grain size (120 nm) in the sample
studied. However, due to rather wide grain-size distribution and
other reasons it cannot be excluded that some grains are
multidomain. It can be seen in Fig. 1 that certain of the
particles are polycrystalline. It follows from Fig. 1 as well that
some of the small particles can stick together rather strongly
and, after stabilization by $\beta$-CrOOH, they can make
multidomain grains. Consequently, the sample studied can be a
mixture of single-domain and multidomain grains.
\par
In an isolated single-domain grain, increase in applied field
above some nucleation field $H_N$ leads to rotation of the
magnetization out from the easy axis into the direction of the
applied field (this mechanism of magnetizing is called homogeneous
rotation) \cite{kron}. For spherical particle $H_N=-2K_{1}/M_{s}$,
where $K_{1}$ is the anisotropy constant. Field $H_N$ determines
upper limit for coercive force $H_c$. $K_1$ is strongly
temperature dependent (reduces with increasing temperature more
rapidly than $M_s$ \cite{morr,chik}), and goes to zero at the
Curie temperature. Typical behavior of $H_{c}(T)$ is suggested to
be the same.
\par
The coercive force $H_c$ in real granular systems is usually much
smaller than $H_N$ for a single particle due to different sources
of inhomogeneities and dependence of $H_c$ on the strength of
intergrain exchange coupling. To include these effects, a general
relation for the coercive fields of real granular ferromagnet is
used \cite{kron}: $H_c=\alpha(2K_{1}/M_s) - N_{eff}M_s,$, where
numerical parameter $\alpha$ ($<1$) takes into account the
microstructure disordering and $N_{eff}$ depends on local
demagnetization fields of grain edges.
\par
The coercive force $H_c$ is a field at which a reversal of the
magnetization takes place when, after magnetizing of a sample for
one field direction, an increasing  magnetic field is applied in
an opposite direction. Magnetic particles of sufficiently large
size will generally not be uniformly magnetized but rather be
composed of magnetic domains. Consequently, in multidomain
particles, the magnetization reverse can be reached not only with
uniform rotation of magnetization but more easily with nucleation
and growth of domain with opposite direction of magnetic moments
\cite{smit,morr,chik}.  For this reason, value of $H_c$ in
multidomain particles is far less than that in single-domain ones.
\par
For homogeneous (single-domain or multidomain) powder systems,
$H_c$ is expected to be maximal at low temperatures but should
decrease with increasing temperature going to zero at $T\simeq
T_C$. The field $H_p$ ($\approx H_c$) follows this expected
behavior, but only above 30 K (Fig. 9). Position of a point of
$H_c$, obtained from magnetization measurement at room
temperature, fits in this behavior (Fig. 9). However, decrease in
$H_p$ with decreasing temperature below 30 K (Fig. 9) is
unexpected in a fairly homogeneous  powder systems and is
evidently closely related with peculiar inhomogeneous structure of
the system studied and percolating character of its conductivity.
In granular systems at low enough temperature there are only a few
conductive current paths. These are "optimal" chains of grains
with maximum conductivity, and it is very likely that they consist
preferably of grains of larger size. Such grains are multidomain
with smaller coercive force $H_c$ that can explain the surprising
behavior of $H_{p}(T)$ in low temperature range (Fig. 9).
\par
In conclusion, we have studied cold-pressed powder of CrO$_2$,
prepared by hydrothermal synthesis, with mean diameter of
particles about 120 nm. The powder was stabilized with thin
surface layer of $\beta$-CrOOH. Transport and magnetotransport
properties of the sample correspond to behavior of magnetic
granular system with weak exchange interaction between the grains.
Large MR in low temperature range can be well ascribed to
spin-dependent intergrain tunnelling. Some peculiar properties of
MR in low-temperature range ($T<40$~K) can be sure attributed to
percolating character of tunnelling conductivity of this granular
systems in conditions of few current paths through the sample.
\par
The authors are indebted to I. L. Potokin for TEM images of the
CrO$_2$ powder.



\vspace{15pt}

\centerline{\bf Figure captions} \vspace{10pt}

Figure 1. Transmission electron micrograph of the CrO$_2$ powder.
\vspace{9pt}

Figure 2. Temperature dependences of magnetization for different
external fields. The dependences were recorded on heating, after
the sample had been cooled down from the room temperature to
$T=77$~K in zero field.\vspace{9pt}

Figure 3. Temperature dependences of the resistivity for different
states of the sample (see main text).\vspace{9pt}

Figure 4. Temperature dependences of the magnetoresistance $\Delta
R(H)/R(0)= [R(H)-R(0)]/R(0)$ at $H=1.2$~T for as-prepared state of
the sample ($\circ$) and after heating the sample up to 430 K
($\bullet$). The inset shows the dependence for the as-prepared
state in semilogarithmic plot, demonstrating exponential
temperature dependence of MR below 200 K.\vspace{9pt}

Figure 5. (Color online) (a) MR curves at $T=5.1$~K recorded with
magnetic field variation $H_{max}\rightarrow -H_{max}$
[MR($\leftarrow$)] and $-H_{max}\rightarrow H_{max}$
[MR($\rightarrow$)], where $H_{max}\approx 1.5$~T. Inset shows
blowup of the MR behavior in low-field range. Peak of positive MR
at field $H_{p}$ is indicated by arrow. (b) Magnetic-field
dependence of difference between curves MR$(\rightarrow)$ and
MR$(\leftarrow)$ taken for opposite directions of the
magnetic-field sweeping.\vspace{9pt}

Figure 6. (Color online) MR curves recorded at $T=7.04$~K.
Measurement protocol is the same as that described in capture to
Fig. 5. Inset shows blowup of the MR behavior in low-field
range.\vspace{9pt}

Figure 7. (Color online) (a) MR curves at $T=15.04$~K recorded
with magnetic field variation $H_{max}\rightarrow -H_{max}$
[MR($\leftarrow$)] and $-H_{max}\rightarrow H_{max}$
[MR($\rightarrow$)]. Inset shows blowup of the MR behavior in
low-field range. (b) Magnetic-field dependence of difference
between the MR$(\rightarrow)$ and MR$(\leftarrow)$ curves taken
for opposite directions of magnetic-field sweeping.\vspace{9pt}

Figure 8. (Color online) MR curves recorded at $T=50$~K. Inset
shows blowup of the MR behavior in low-field range.\vspace{9pt}

Figure 9. Temperature dependence of magnetic field $H_p$ at which
resistance is maximum in MR curves. Symbol $\blacktriangle$
indicates the position of coercive force $H_c$ at room
temperature, obtained in magnetization measurements. The data are
related to as-prepared state of the sample studied.

\newpage
\begin{figure}[tb]
\centering\includegraphics[width=0.7\linewidth]{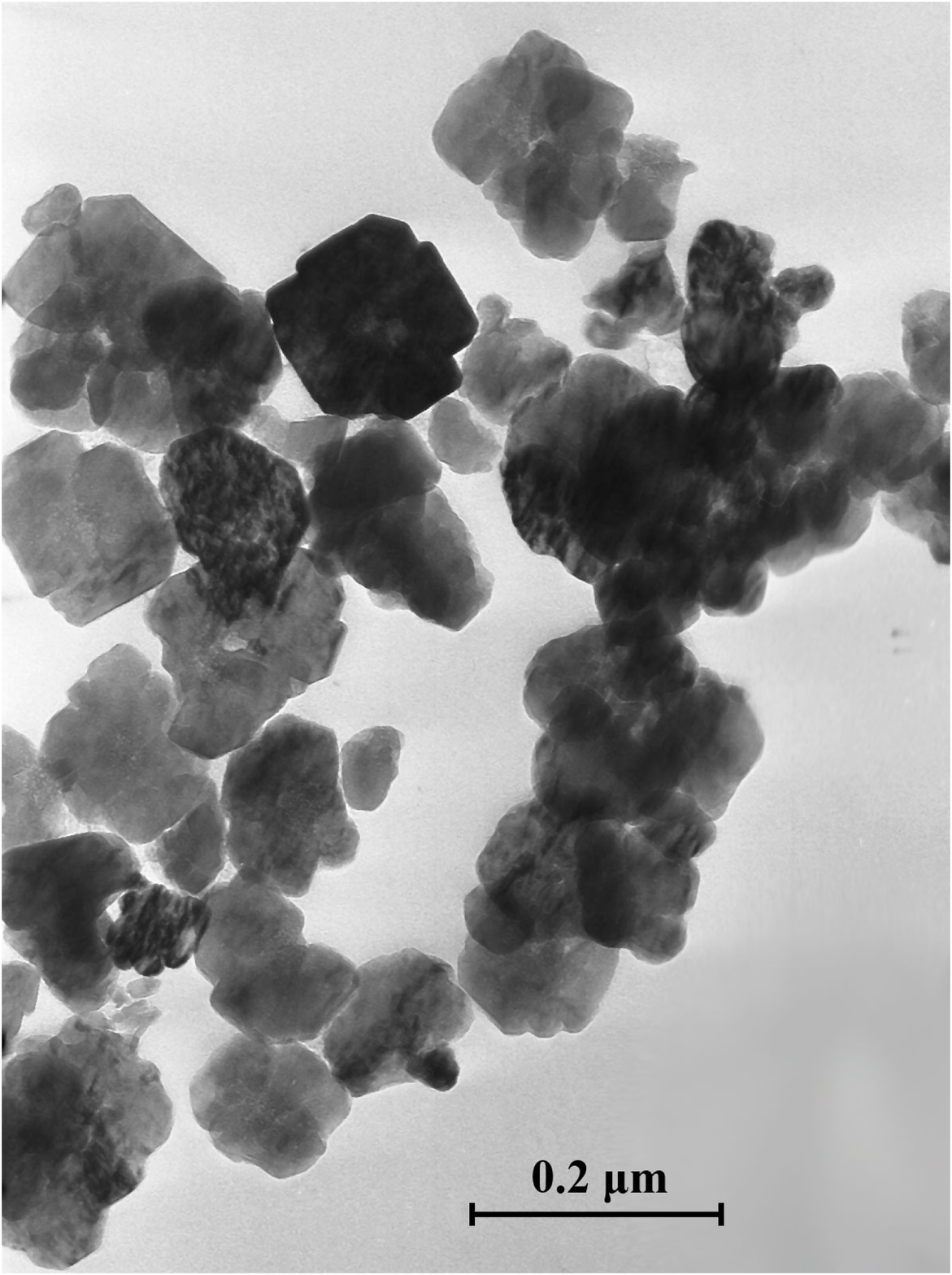}
\vspace{29pt}
\par \centerline{Figure 1 to paper Belevtsev et al.}
\end{figure}

\begin{figure}[tb]
\centering\includegraphics[width=0.85\linewidth]{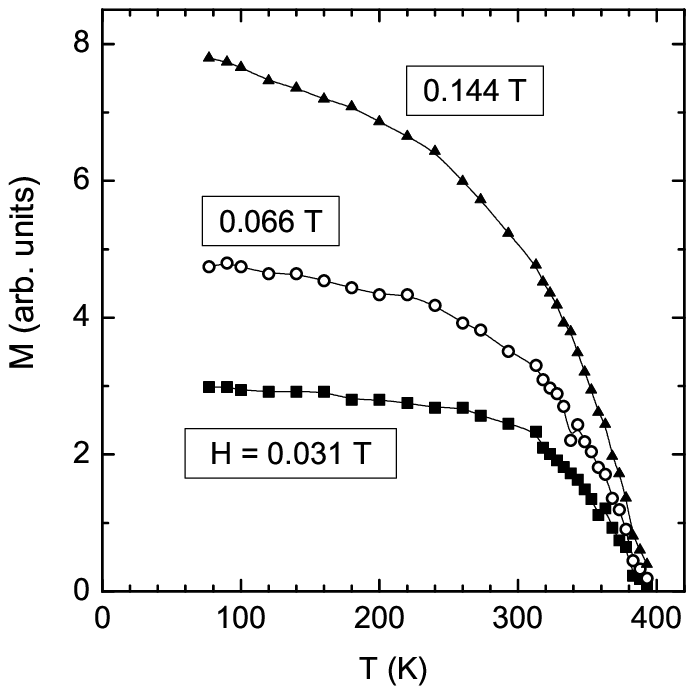}
\vspace{29pt}
\par \centerline{Figure 2 to paper Belevtsev et al.}
\end{figure}

\begin{figure}[tb]
\centering\includegraphics[width=0.85\linewidth]{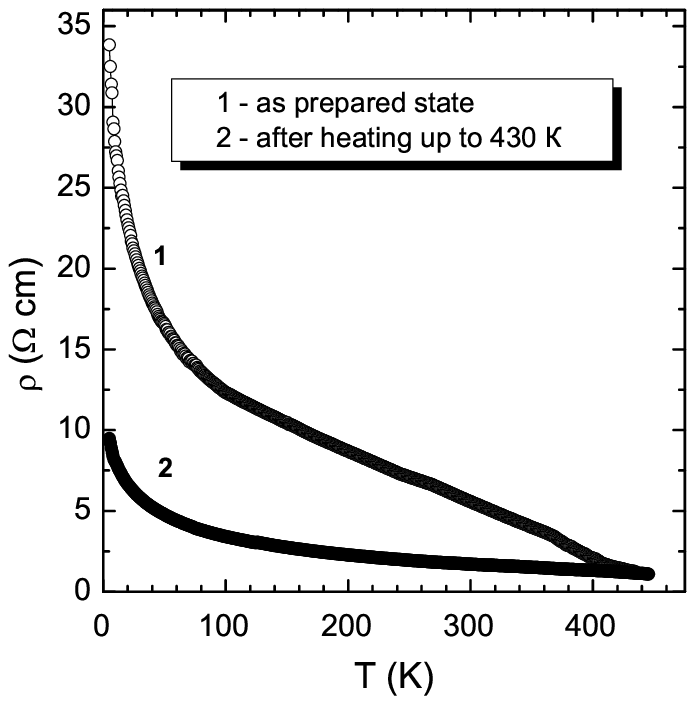}
\vspace{29pt}
\par \centerline{Figure 3 to paper Belevtsev et al.}
\end{figure}

\begin{figure}[tb]
\centering\includegraphics[width=0.85\linewidth]{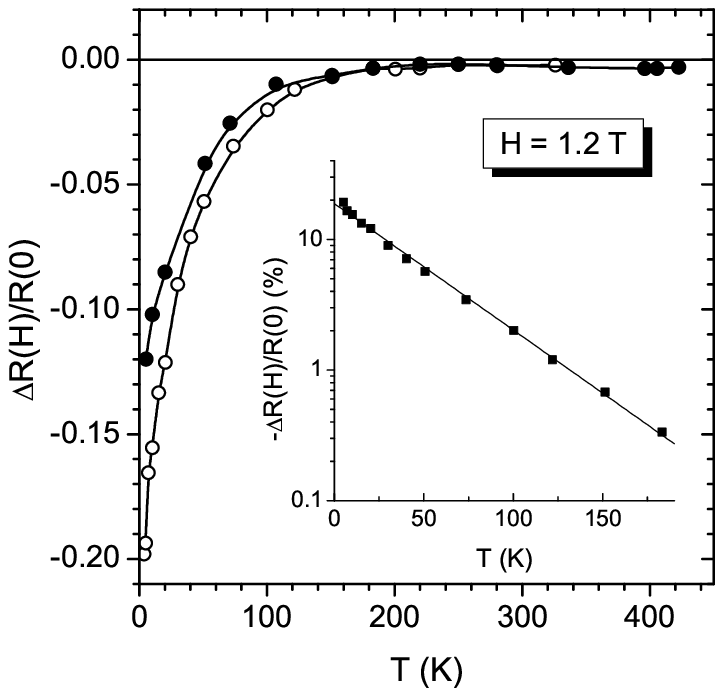}
\vspace{29pt}
\par \centerline{Figure 4 to paper Belevtsev et al.}
\end{figure}

\begin{figure}[tb]
\centering\includegraphics[width=0.85\linewidth]{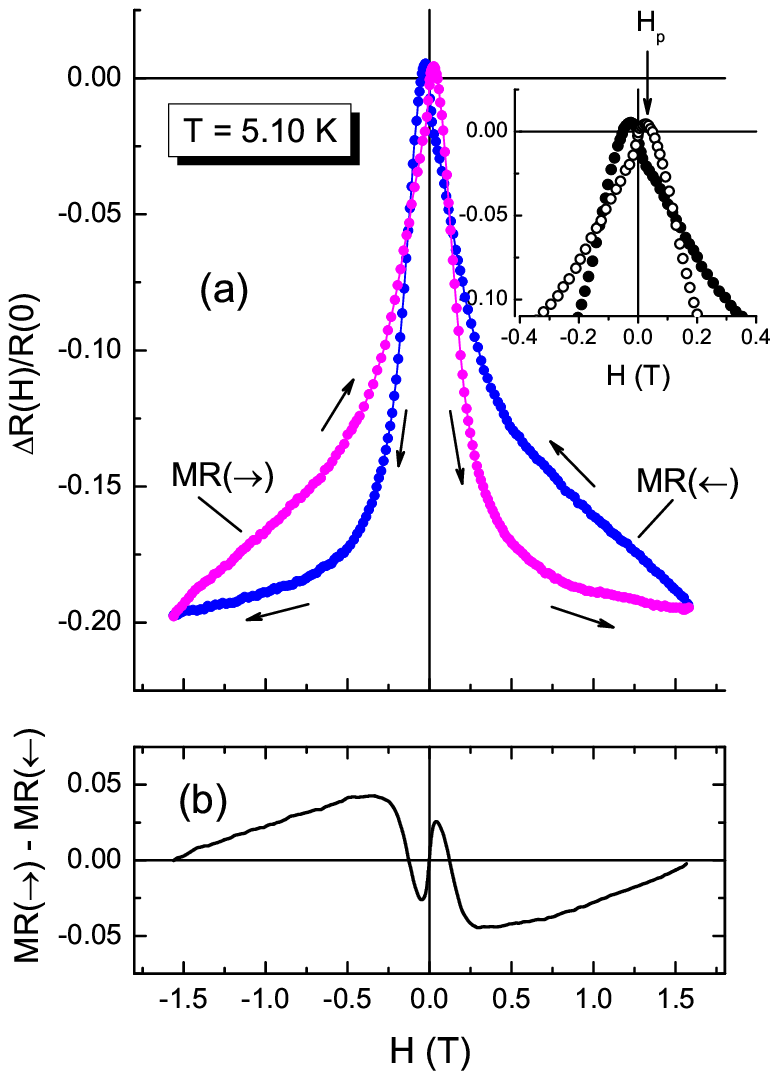}
\vspace{29pt}
\par \centerline{Figure 5 to paper Belevtsev et al.}
\end{figure}

\begin{figure}[tb]
\centering\includegraphics[width=0.85\linewidth]{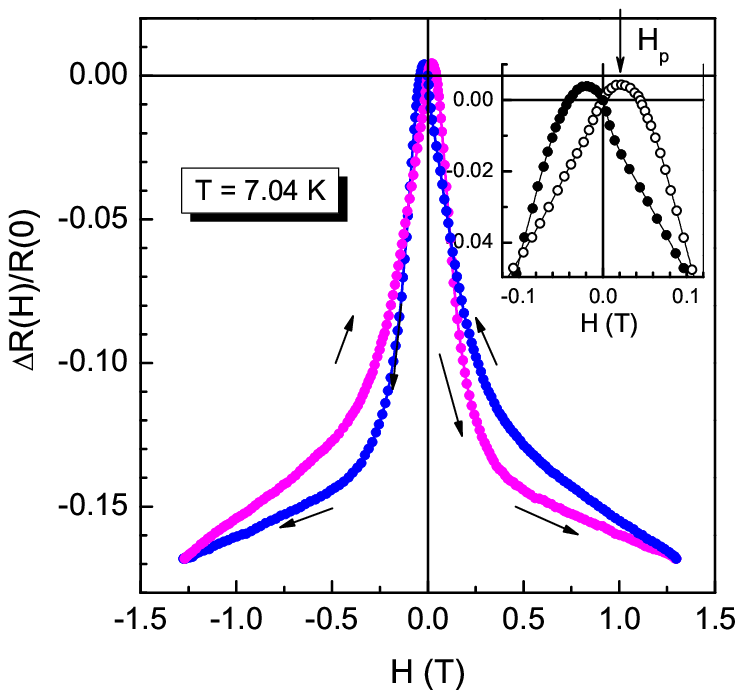}
\vspace{29pt}
\par \centerline{Figure 6 to paper Belevtsev et al.}
\end{figure}

\begin{figure}[tb]
\centering\includegraphics[width=0.85\linewidth]{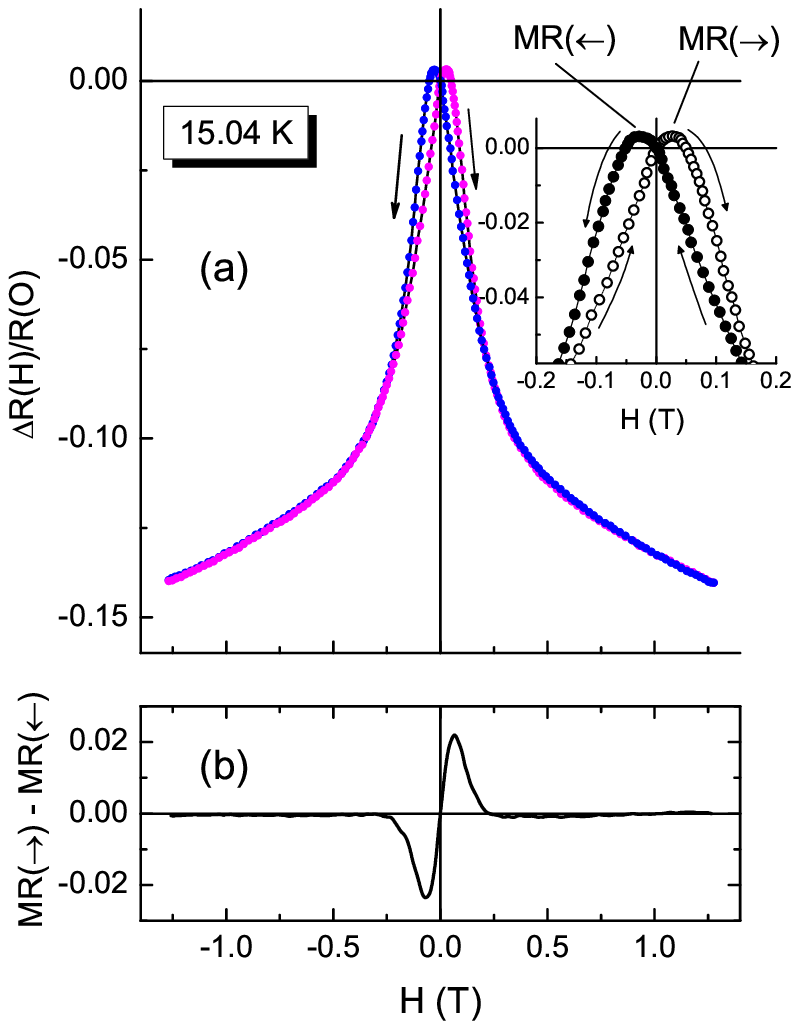}
\vspace{29pt}
\par \centerline{Figure 7 to paper Belevtsev et al.}
\end{figure}

\begin{figure}[tb]
\centering\includegraphics[width=0.85\linewidth]{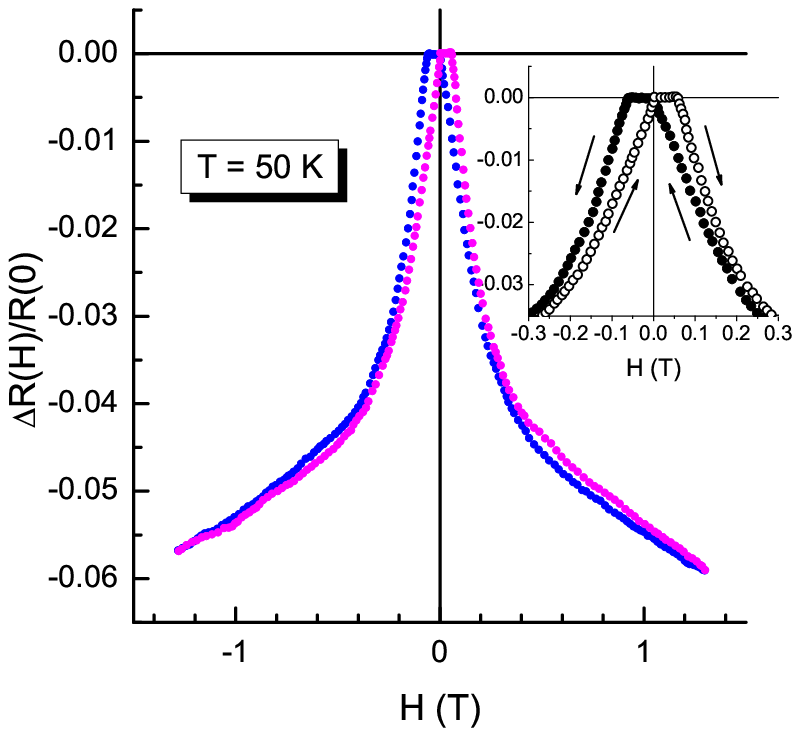}
\vspace{29pt}
\par \centerline{Figure 8 to paper Belevtsev et al.}
\end{figure}

\begin{figure}[tb]
\centering\includegraphics[width=0.85\linewidth]{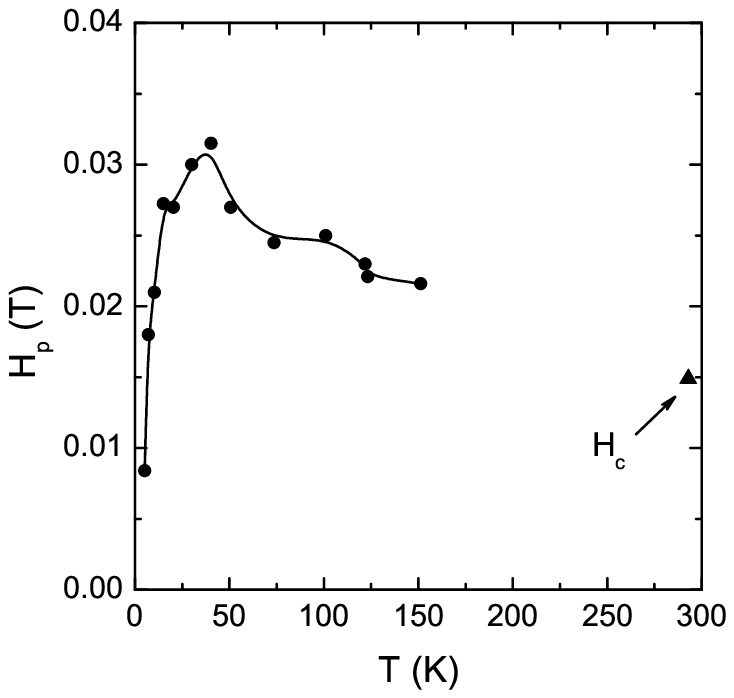}
\vspace{29pt}
\par \centerline{Figure 9 to paper Belevtsev et al.}
\end{figure}

\end{document}